\documentclass[twocolumn,twoside,slac_two]{revtex4}
\usepackage{graphicx}
\usepackage{fancyhdr}
\input babarsym

\newcommand{\pppz}{\ensuremath{\pim\pip\piz}}
\newcommand{\kppz}{\ensuremath{K^-\pip\piz}}
\newcommand{\kkpz}{\ensuremath{K^-K^+\piz}}

\def\Mnu{\ensuremath{{\cal M}_\nu^2}}
\def\psoft {\ensuremath{\pi^+_s}\xspace}

\newcommand{\dss} {\ensuremath{D^{*+}_s}}
\newcommand{\dso} {\ensuremath{D^-_{s1}}}
\newcommand{\ds}  {\ensuremath{D^+_s}}
\newcommand{\ks}  {\ensuremath{K_S^0}}
\newcommand{\dstbkm} {\ensuremath{\overline{D}{}^{*0}K^-}}
\newcommand{\dstmks} {\ensuremath{D^{*-}\ks}}

\newcommand{\Dzpp}{\ensuremath{\Dz \to \pim\pip}}
\newcommand{\Dzkk}{\ensuremath{\Dz \to K^{-}K^{+}}}
\newcommand{\Dzpppz}{\ensuremath{\Dz \to \pim\pip\piz}}
\newcommand{\kmkppz}{\ensuremath{K^- K^+ \pi^0}}

\newcommand{\DstarDzpis}{\ensuremath{\Dstarp \to \Dz \pip_s}}
\newcommand{\Dzkppz}{\ensuremath{\Dz \to K^-\pip\piz}}
\newcommand{\Dzkkpz}{\ensuremath{\Dz \to K^{-}K^{+}\piz}}
\newcommand{\RM}  {\ensuremath{ M_{\mathrm{recoil}} }}
\newcommand{\RMD} {\ensuremath{\Delta M_{\mathrm{recoil}} }}

\def\piz   {\ensuremath{\pi^0}\xspace}
\def\pip   {\ensuremath{\pi^+}\xspace}
\def\Dztilde   {\ensuremath {\tilde{D}^0}\xspace}

\begin{document}

\title{$\mathbf{D}$ and $\mathbf{D_s}$ hadronic branching fractions at B factories}

%

\author{M. Pappagallo(on behalf of the \babar\  Collaboration)}
\affiliation{University of Bari and I.N.F.N., 70126 Bari, Italy}

\begin{abstract}
Recent measurements of hadronic branching fractions of $D$ and $D_s$ mesons, performed by the \babar\ and Belle experiments at the asymmetric $e^+e^-$ B factories colliders PEP II and KEKB, are reviewed.
\end{abstract}

\maketitle

\thispagestyle{fancy}


\section{Introduction}

Hadronic branching fractions of $D$ and $D_s$ decays are used as
references mode in many measurements of branching fractions of $D$ and $B$-meson decays as well. A precise measurement of such values improves our knowledge of $D$ and $B$-meson properties, and of fundamental parameters of the Standard Model, such as the magnitude of the Cabibbo-Kobayashi-Maskawa~\cite{CKM} matrix element.
\section{Absolute branching fraction of $\mathbf{D^0 \to K^- \pi^+}$}
\babar\ collaboration measures the absolute branching fraction ${\cal B}(\Dz \rightarrow \Km \pip)$\footnote{Charge conjugation is implied through the paper.} using  $\Dz \rightarrow \Km \pip$ decays in a sample of $\Dz$ mesons preselected
by their production in \dsp\ decays, obtained with partial reconstruction of the decay
 $\Bzb \rightarrow D^{*+} X \ell^{-} \bar{\nu}_{\ell}$, with $\dsp
\rightarrow \Dz \pip$~\cite{Aubert:2007wn}. Such measurement is extremely important because many of the past and current $D$ and $B$ branching fraction measurements are indeed systematically limited by the precision of ${\cal B}(\Dz \rightarrow \Km \pip)$.

A sample of partially reconstructed $B$ mesons in the channel
$\Bzb \rightarrow D^{*+} X \ell^{-} \bar{\nu}_{\ell}$ is selected
by retaining events containing a charged lepton ($\ell = e,\,\mu$) and a low momentum
pion (soft pion, $\pi^+_{s}$) which may arise from the decay $D^{*+}\to \Dz \pi^+_{s}$.
This sample of events is referred to as the ``inclusive sample''.

\begin{figure}[htb]
\centering
\includegraphics[width=80mm]{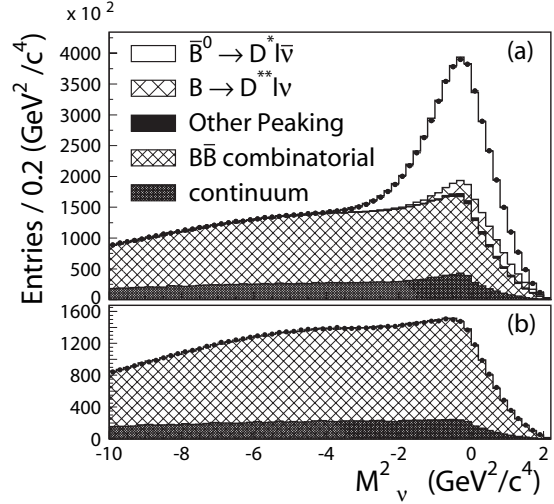}
\caption{The ${\cal M}_\nu^{\,2}$ distribution of the inclusive
  sample, for right-charge (a) and wrong-charge (b) samples.
The data are represented by solid points with error.
The MC fit results are overlaid to the data, as explained in the figure.}
\label{fig:incl_yield}
\end{figure}

Using conservation of momentum and energy,
the invariant mass squared of the undetected neutrino is calculated as
$$\Mnu \equiv (E_{\mbox{\rm \small beam}}-E_{{D^*}} -
E_{\ell})^2-({\vec{p}}_{{D^*}} + {\vec{p}}_{\ell})^2 ,$$
where $E_{\mbox{\rm \small beam}}$ is half the total center-of-mass energy, $E_{\ell}~(E_{{D^*}})$,
 ${\vec{p}}_{\ell}~({\vec{p}}_{{D^*}})$ are the energy and momentum
of the lepton (the $D^*$ meson) and the magnitude of the $B$ meson
momentum, $p_{B}$, is considered negligible compared to $p_{\ell}$ and $p_{D^*}$.
Figure~\ref{fig:incl_yield} shows the \Mnu\ distribution and the results of  a  minimum $\chi^2$ fit aiming to determine the signal and background contribution.  The number of signal events with $\Mnu > -2$ GeV$^2$/c$^4$ results $N^{\rm incl} = (2170.64 \pm 3.04 (\textrm{stat}) \pm 18.1 (\textrm{syst})) \times 10^3$.

The $\Dz \to K^- \pip$ decays in the inclusive sample are selected requiring events in the mass range $1.82 < M_{K\pi} < 1.91$ \gevcc and $142.4 < \Delta M < 149.9$ \mevcc where $\Delta M = M(\Km \pip \psoft) - M(\Km \pip)$ and $\psoft$ is the slow pion from $D^{*+}$ decay. The exclusive selection yields $N^{\rm excl} = 33810 \pm 290$ signal events, where the error is statistical only.

The branching fraction is computed as
$${\cal B}(\Dz \rightarrow \Km \pip) = {N^{\rm excl} \over N^{\rm incl}} \frac{1}{\varepsilon_{(\Km \pip)}\zeta},$$
where $\varepsilon_{(\Km \pip)}$ is the $\Dz$
reconstruction efficiency as computed in the simulation, and
$\zeta$ is the selection bias introduced by the
partial reconstruction.

The main systematic uncertainty on $N^{\rm incl}$ and $N^{\rm excl}$ are respectively due to
the non-peaking combinatorial \BB\ background and the charged-track reconstruction efficiency.
The complete set of systematic uncertainties is listed in Tab.~\ref{tab:sys_errors}.
The absolute branching fraction of $\Dz \rightarrow \Km \pip$ decay
results
$${\cal B}(\Dz \rightarrow \Km \pip) = (4.007 \pm 0.037 \pm 0.070)\% ,$$
where the first error is statistical and the second error is
systematic. This result is comparable in precision with
the present world average, and it is consistent with it within two
standard deviations.
\begin{table}[!h]
\caption{The relative systematic errors of ${\cal B}(\Dz \rightarrow \Km \pip)$.}
\begin{center}
\begin{tabular}{@{}ll@{}c@{}} \hline \hline
                       &Source                                &$\delta({\cal B})/{\cal B}$(\%)\\ \hline
                       &Selection bias                        &$\pm 0.35$        \\  \hline
$N^{\rm incl}$          &Non-peaking combinatorial background  &$\pm 0.89$        \\
                       &Peaking combinatorial background      &$\pm 0.34$        \\
                       &Soft pion decays in flight            &$\pm 0.10$        \\
                       &Fake leptons                          &$\pm 0.08$        \\
                       &Cascade decays                        &$\pm 0.08$        \\
                       &Monte Carlo events shape              &$\pm 0.08$        \\
                       &Continuum background                  &$\pm 0.05$        \\
                       &\dstrstr\ production                  &$\pm 0.02$        \\
                       &Photon radiation                      &$\pm 0.02$        \\  \hline
$N^{\rm excl}$          &Tracking efficiency                   &$\pm 1.00$        \\
                       &$K^-$ identification                  &$\pm 0.70$        \\
                       &\Dz invariant mass                    &$\pm 0.56$        \\
                       &Combinatorial background shape        &$\pm 0.30$        \\
                       &Combinatorial background normalization&$\pm 0.16$        \\
                       &Soft pion decay                       &$\pm 0.12$        \\
                       &Cabibbo-suppressed decays             &$\pm 0.10$        \\
                       &Photon radiation in \Dz\ decay        &$\pm 0.07$        \\  \hline
Total                  &                                      &$\pm 1.74$        \\  \hline \hline
\end{tabular}
\end{center}
\label{tab:sys_errors}
\end{table}

\section{Absolute branching fraction of $\mathbf{D_s^+ \to K^+ K^- \pi^+}$}
\begin{figure*}[t!]
\centering
\includegraphics[width=80mm]{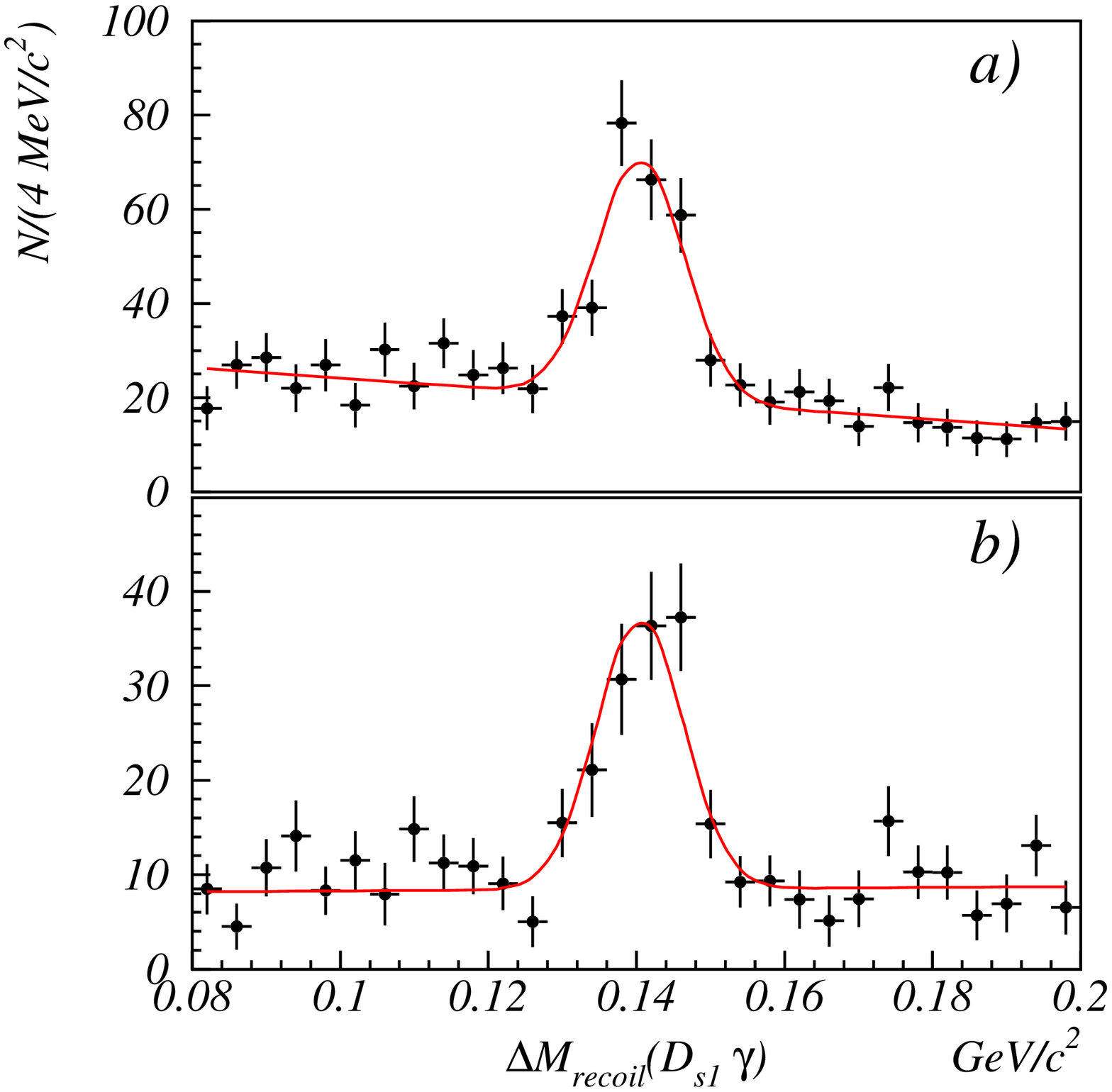}
\includegraphics[width=80mm]{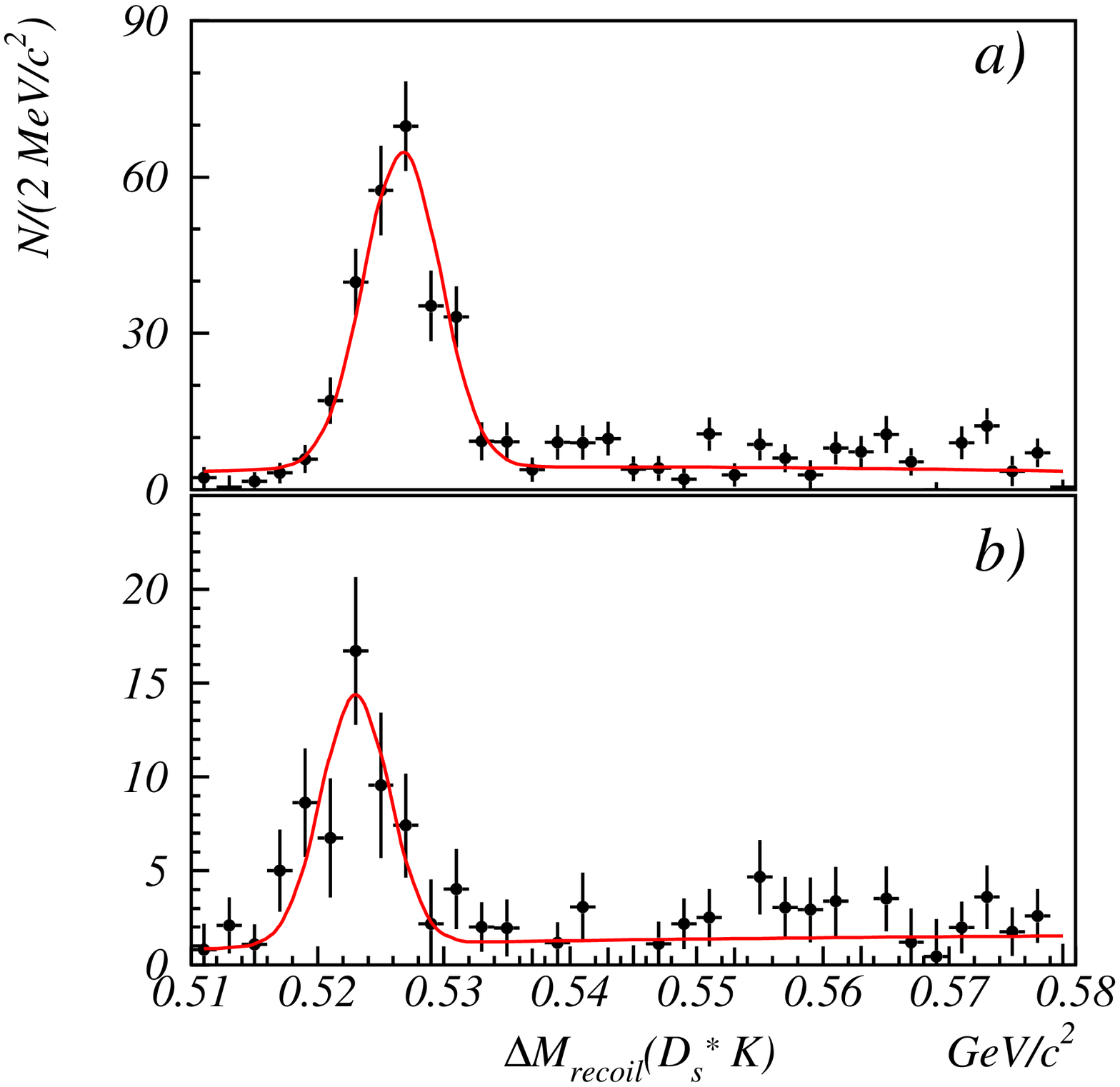}
\caption{The $\ee\to\dss\dso$ signal yields in bins of $\RMD(\dso\gamma)$({\it left}) and $\RMD(\dss K)$({\it right}):
a) for the $\dso \to \dstbkm $ channel and b) for the $\dso \to \dstmks$ channel.
}
\label{fig:ds_kkpi}
\end{figure*}
The  poor accuracy of the branching fraction ${\cal B}(D_s^+ \to K^+ K^- \pi^+)=(5.2\pm 0.9)\%$~\cite{PDG} has been  a systematic limitation for some precise measurements. In particular, the recent study of the $CP$ violation in $B^0 \to D^{(*)\pm} \pi^{\mp}$ decays is restricted by the knowledge of the ratio of two amplitudes that determine  the $CP$-asymmetry~\cite{belle_dp,babar_dp}. The amplitude $B^0 \to D^{(*)+} \pi^-$ can be calculated from the branching fraction of $B^0 \to D_s^{(*)+} \pi^-$ decays assuming  factorization. On the other hand, the factorization hypothesis can be tested by measuring the ratio of  $B^0 \to D^{(*)-} \pi^+$ and $B^0 \to D^{(*)-} D_s^+$ decays. Both  $\mathcal{B}(B^0 \to D_s^{(*)+} \pi^-)$ and $\mathcal{B}(B^0 \to D^{(*)-} D_s^+)$ measurements can be improved with better accuracy in $D_s^+$ absolute branching fractions.

Belle collaboration measures ${\cal B}(D_s^+ \to K^+ K^- \pi^+)$ using a partial reconstruction of the
process $\ee \to \dss D_{s1}^-$~\cite{Abe:2007jz}. In this analysis 4-momentum
conservation allows to infer the 4-momentum of the undetected
part.

The process $\ee \to \dss D_{s1}^-$ is reconstructed using two
different tagging procedures. The first one (denoted as the \dso\ tag)
includes the full reconstruction of the \dso\ meson via $\dso\to
\overline{D}{}^* K$ decay and observation of the photon from $\dss \to
\ds \gamma$, while the $D_s^+$ is not reconstructed. The measured
signal yield with the \dso\ tag is proportional to the branching fractions
of the reconstructed $\overline{D}{}^*$ modes. In the second procedure
(denoted as the \dss\ tag) a full reconstruction of \dss\
is required through $\dss \to \ds \gamma$ and observation of the kaon from
$D_{s1}^{-}\to \overline{D}{}^{*} K$, but the $\overline{D}{}^{*}$ is
not reconstructed. Since the \ds\ meson is reconstructed in the
channel of interest, $\ds\to K^+ K^- \pi^+$, the signal yield measured
with the \dss\ tag is proportional to this \ds\ branching
fraction. The (efficiency-corrected) ratio of the two measured signal
yields is equal to the ratio of well-known $\overline{D}{}^{*}$
branching fractions and the branching fraction of the \ds:
\begin{eqnarray}
  \mathcal{B} (\ds \to K^+ K^- \pi^+ )= \frac{N(\dss)}
  {N(\dso)}\cdot \frac{ \epsilon(\dso)}{\epsilon
  (\dss)} \mathcal{B}(\overline{D}{}^{(*)}),
\label{eq:ds_kkpi}
\end{eqnarray}
where $\mathcal{B}(\overline{D}{}^{(*)})$ is the product of
$\overline{D}{}^*$ branching fraction and those of sub-decays.

The signal is identified by studying the mass recoiling against the
reconstructed particle (or combination of particles) denoted as $X$. This recoil mass is defined as:
$$\RM(X) \equiv \sqrt{ {(E_{CM} - E_X)}^2 - P_X^2 },$$
where $E_X$ and $P_X$ are the center-of-mass (CM) energy and momentum of X,
respectively; $E_{CM}$ is the CM beam energy. A peak in the \RM\ distribution at the
nominal mass of the recoil particle is expected.

Since the resolution in \RM\ is not enough to separate the relevant final states, the recoil mass difference \RMD\ is used to
disentangle the  contribution of the different final states:
$$\RMD(D^{-}_{s1} \gamma) \equiv \RM(D^{-}_{s1})-\RM(D^{-}_{s1}\gamma),$$
$$\RMD(D^{*+}_s K) \equiv \RM(D^{*+}_s)-\RM(D^{*+}_s K).$$

As the ratio of $\dso\to\dstbkm$  and $\dso\to
\dstmks$ branching fractions is unknown, the analysis is performed for
these two channels separately. Figure~\ref{fig:ds_kkpi} shows the $\RMD(\dso\gamma)$ and $\RMD(\dss K)$ distributions used for $D_{s1}^-$ and $D_s^{*+}$ tag procedures respectively. $\RMD(\dso\gamma)$ peaks at around $\simeq 0.14$ GeV/c$^2 \simeq M(D_s^*)-M(D_s)$. $\RMD(\dss K)$ peaks at around $\simeq 0.525$ GeV/c$^2 \simeq M(D_{s1})-M(D^*)$.

Using the measured signal yields $N(\dss)$ and $N(\dso)$ with \dss\
and \dso\ tags, respectively, and taking into account the
efficiency ratio $\frac{ \epsilon(\dso)}{\epsilon(\dss)}$, the \ds\ absolute
branching fraction is computed by Eq.~\ref{eq:ds_kkpi} for $\dso\to\dstbkm$  and $\dso\to \dstmks$. The average value is $\mathcal{B}(\ds \to K^+ K^-
\pi^+)=(\ensuremath{ 4.0 \pm 0.4(\textrm{stat}) \pm 0.4(\textrm{syst})}$)\%.

\section{Relative branching fraction of $\mathbf{D^0 \to K^-K^-\pi^0}$ and $\mathbf{D^0 \to \pi^+ \pi^- \pi^0}$}

The branching ratios of the singly Cabibbo-suppressed decays of \Dz\ meson
are anomalous since the \Dzpp\ branching fraction is observed to be suppressed
relative to the \Dzkk\ by a factor of almost three, even though the phase
space for the former is larger. The branching ratios of the three-body decays~\cite{PDG} have larger uncertainties but do not appear to exhibit the same suppression. This motivates the current study which measures the branching ratios of \Dzpppz\ and \kmkppz\ with respect to the Cabbibo-favored decay \Dzkppz. \babar\ collaboration measures both branching ratios~\cite{Aubert:2006xw}, Belle collaboration only the decay \Dzpppz\ with respect to the decay \Dzkppz~\cite{Abe:2006tv}.
By choosing the normalization mode $D^0 \to K^-\pi^+\pi^0$, many
sources of systematic uncertainty including the $\pi^0$ detection efficiency
and uncertainty in the tracking efficiency cancel out.
To reduce combinatorial backgrounds, \Dz\ candidates are reconstructed in decays
$\DstarDzpis$ ($\pi_s^+$ is a soft, low momentum charged pion) with $\Dzkppz$, $\pppz$,
and $\kkpz$, by selecting events with at least three charged tracks and a neutral pion.

\begin{figure*}[t!]
\centering
\includegraphics[width=55mm]{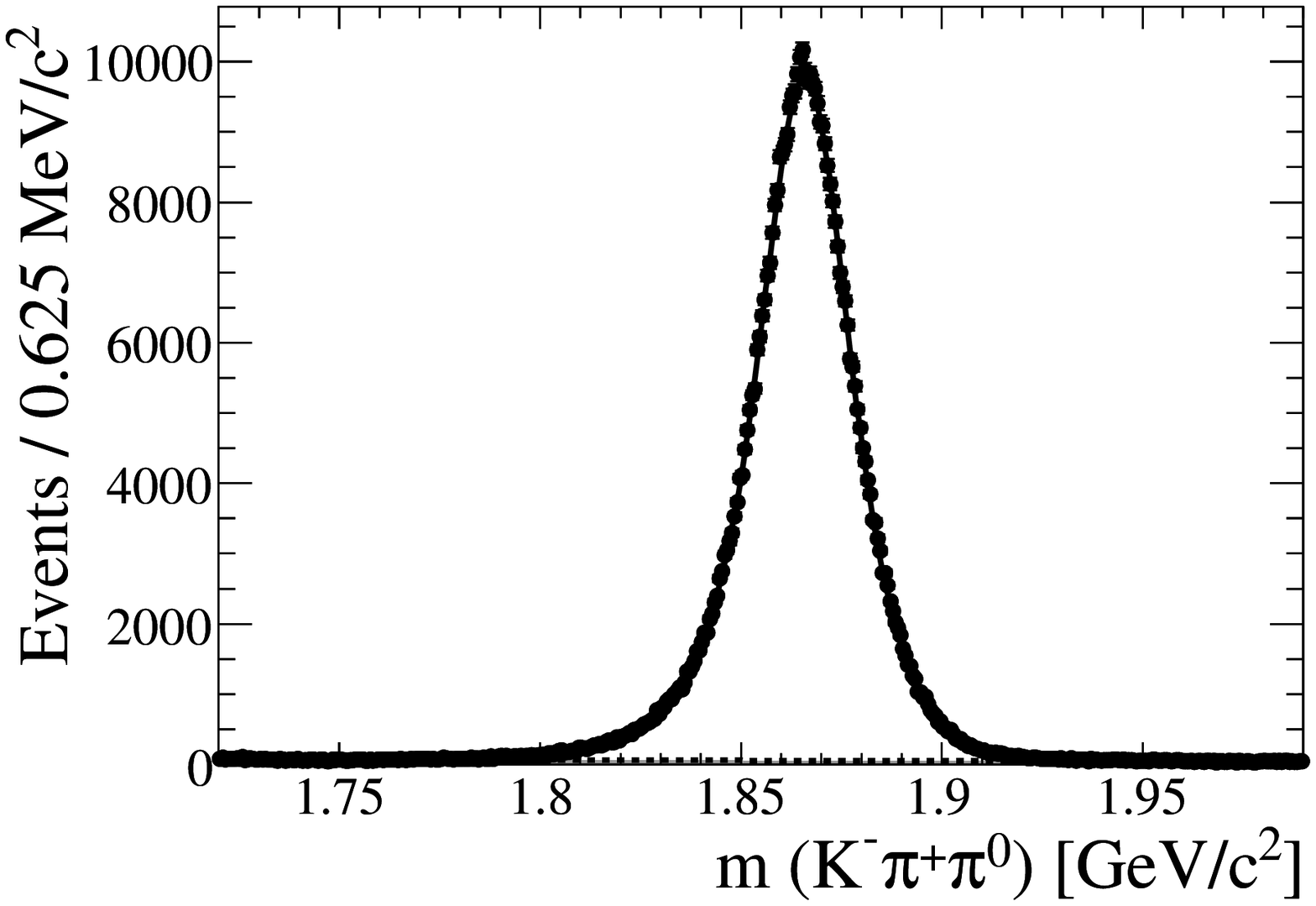}
\includegraphics[width=55mm]{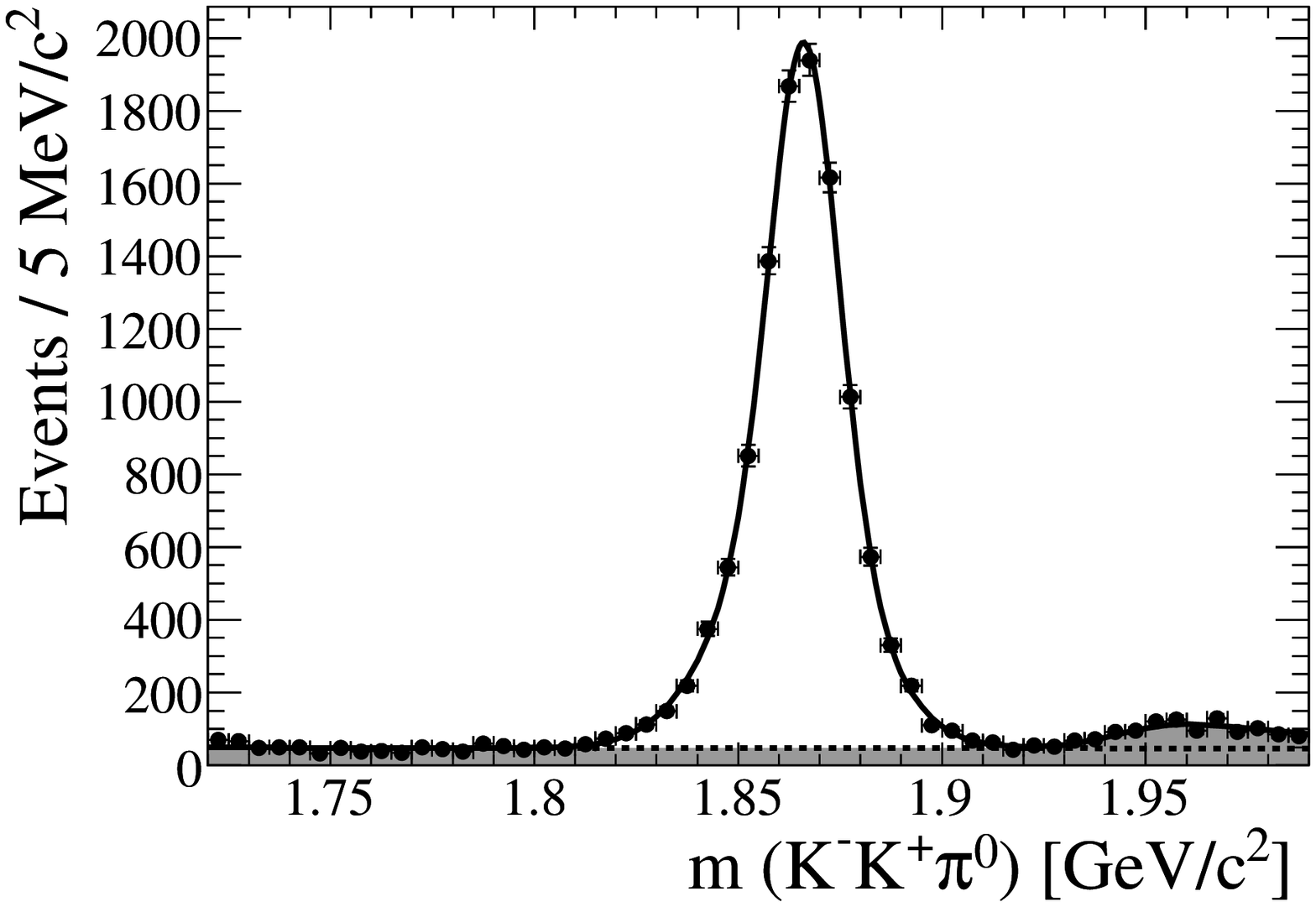}
\includegraphics[width=55mm]{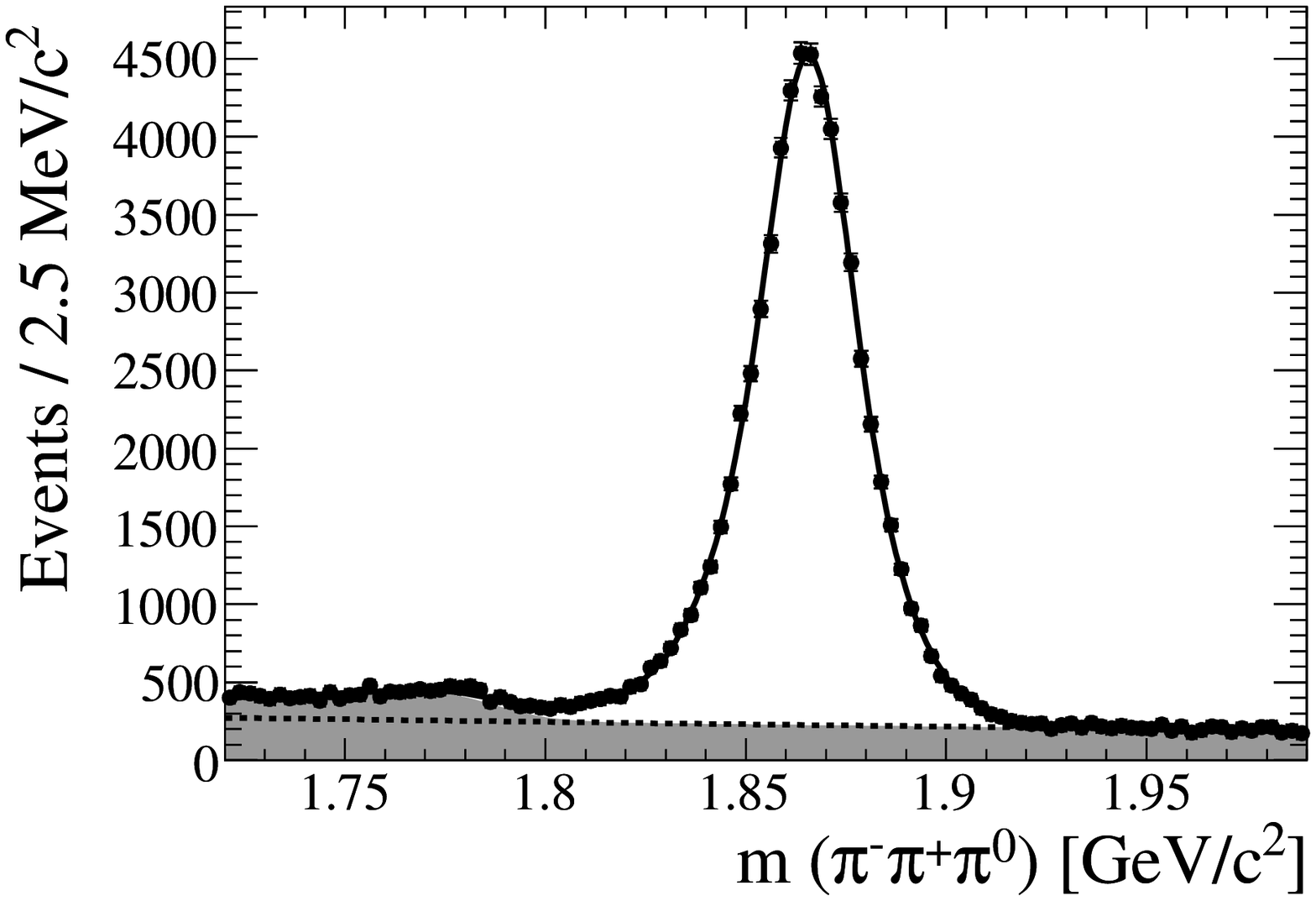}
\includegraphics[width=45mm]{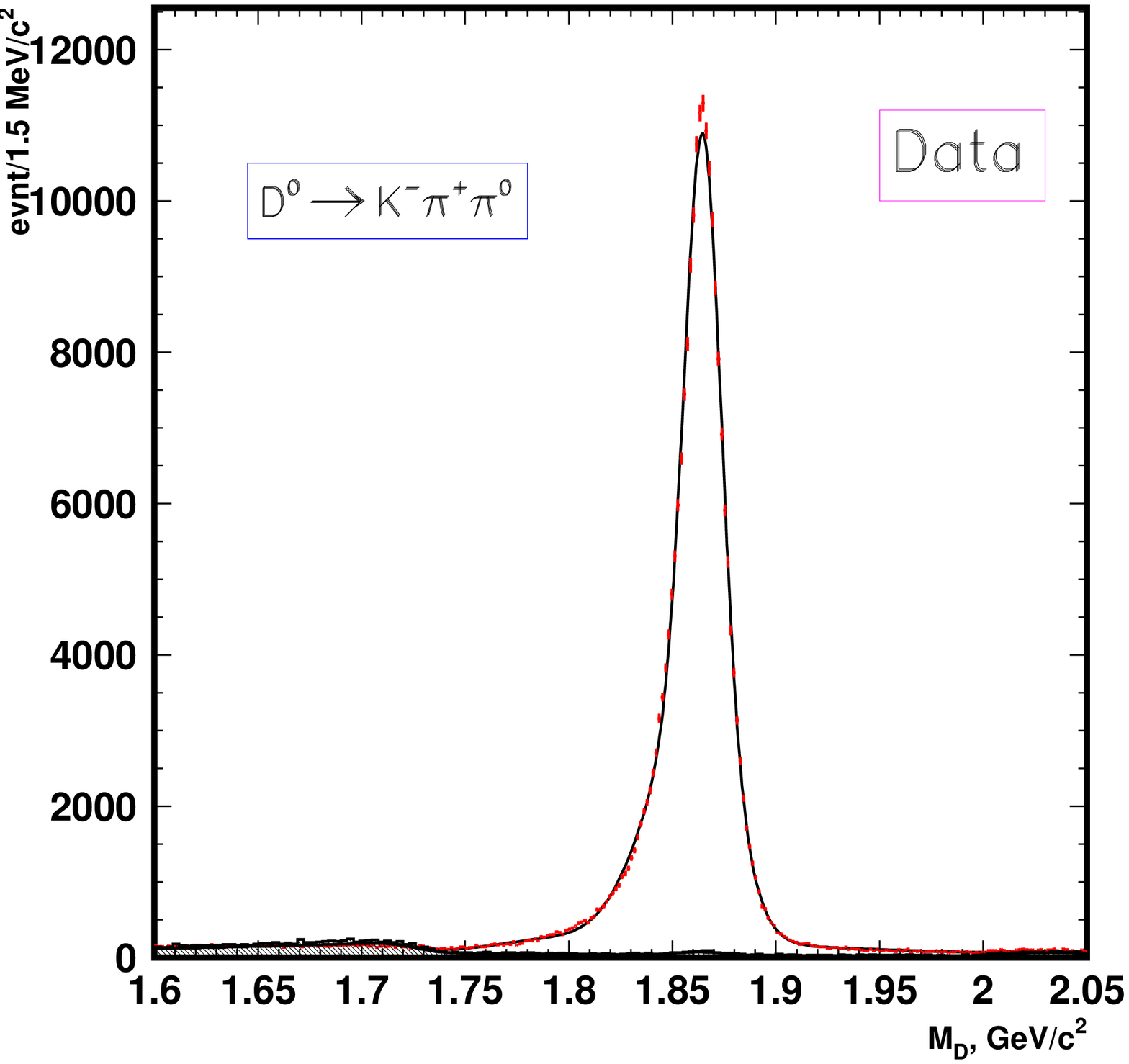}
\includegraphics[width=45mm]{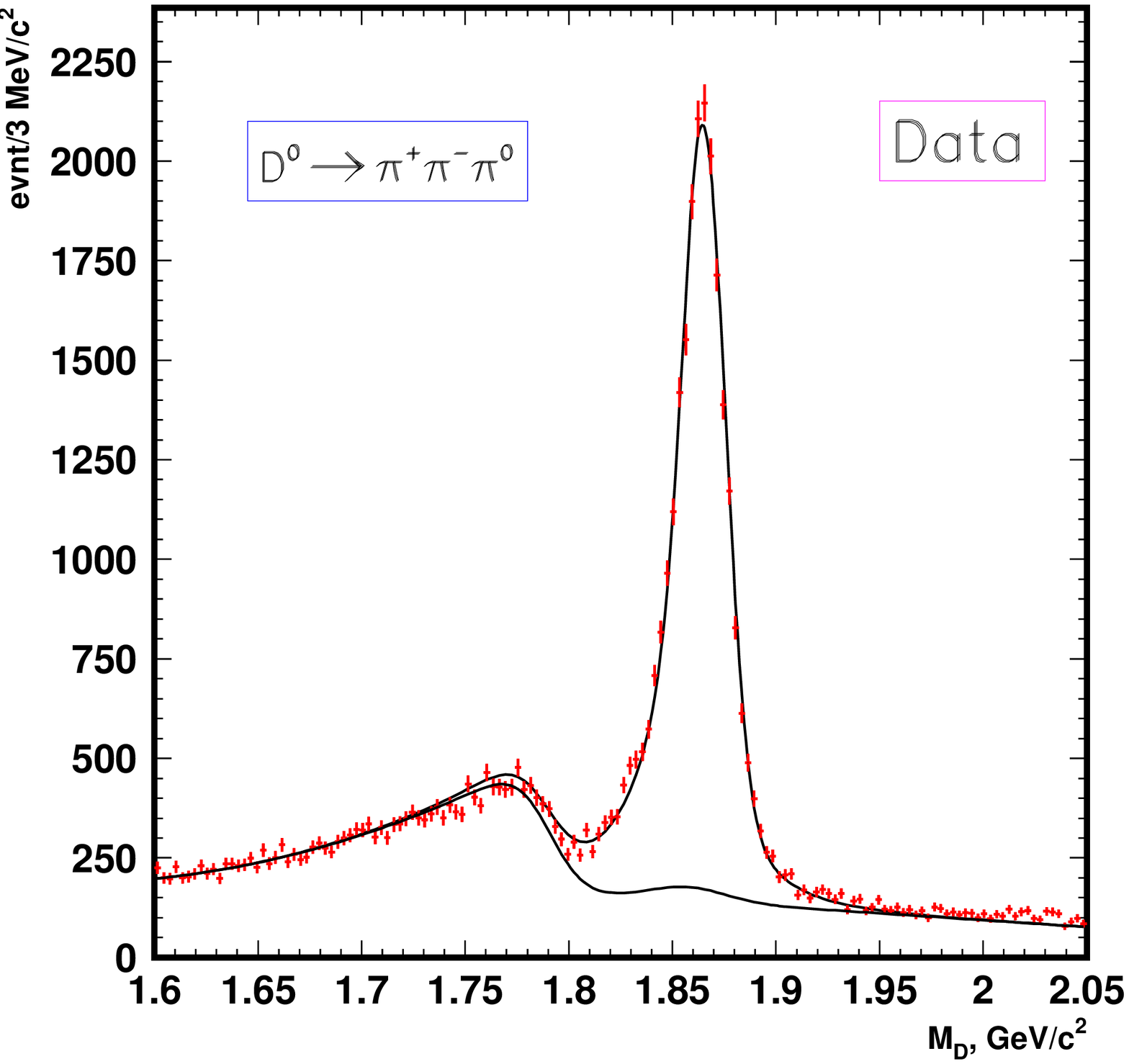}


\caption{{\it Top}(\babar\ collaboration): Fitted mass for the \kppz, \pppz, and \kkpz\ data samples. Dots are data
points and the solid curves are the fit. The dot-dashed
lines show the level of combinatorial background in each case. For the \pppz\ and
the \kkpz\ modes, the shaded region represents the total background. {\it Bottom}(Belle collaboration): Signal $M(K\pi\pi^0)$ distribution fitted with 2 bifurcated
Gaussians + Gaussian (signal peak) and the generic MC shape (background). Signal $M(\pi^+\pi^-\pi^0)$ distribution, fitted
 to the signal MC shape for the signal peak and with the generic MC shape
 (background).
 }

\label{fig:BR:pipipi0}
\end{figure*}

\babar\ obtains the following results for the branching ratios:
%
$$ \frac{{\cal B}(\Dzpppz)}{{\cal B}(\Dzkppz)} = ( 10.59 \pm 0.06 \pm 0.13 ) \times 10^{-2},$$
$$ \frac{{\cal B}(\Dzkkpz)}{{\cal B}(\Dzkppz)} = ( 2.37 \pm 0.03 \pm 0.04 ) \times 10^{-2},$$
%
while Belle obtains:
$$\frac{{\cal B}(D^0 \to \pi^+\pi^-\pi^0)}
{{\cal B}(D^0 \to K^-
\pi^+\pi^0)}= (9.71 \pm 0.09 \pm 0.30)\times 10^{-2}. $$\\
Errors are statistical and systematic, respectively.
Figure~\ref{fig:BR:pipipi0} shows the resulting mass distributions. Reflected \kppz\ events peak in the lower (upper) sideband of $m_{\pppz}$ ($m_{\kkpz}$).

Using the world average value for the $\Dzkppz$ branching fraction~\cite{PDG}, the absolute branching ratios result:\\

\noindent \babar
%
$${\cal B}(\Dzpppz) = ( 1.493 \pm 0.008 \pm 0.018 \pm 0.053 ) \%,$$
$${\cal B}(\Dzkkpz) = ( 0.334 \pm 0.004 \pm 0.006 \pm 0.012 ) \%,$$
%
Belle
%
$${\cal B}(\Dzpppz) = ( 1.369 \pm 0.013 \pm 0.042 \pm 0.049 ) \%,$$
%
where the errors are statistical, systematic, and due to the uncertainty of
${\cal B}(\Dzkppz)$. \\
\indent The decay rate for each process can be written as:
$$\Gamma = \int d\Phi |{\cal M}|^2,$$
\noindent where $\Gamma$ is the decay rate to a particular three-body final state,
${\cal M}$ is the decay matrix element, and $ \Phi $ is the phase space.
Integrating over the Dalitz plot assuming a uniform phase space density, the above
equation can be written as:
$$\Gamma = \langle | {\cal M} | ^2 \rangle \ \times \Phi,$$
where $ \langle | {\cal M} | ^2 \rangle $ is the average
value of $ | {\cal M} | ^2 $ over the Dalitz plot and the
three-body phase space, $\Phi$ is proportional to the area of the Dalitz plot.
For the three signal decays $\Phi$  is in the ratio $\pppz:\kppz:\kkpz$  = 5.05 : 3.19 : 1.67.
Combining the statistical and systematic errors, it results:\\

\noindent \babar
\begin{equation}
{ {\langle|{\cal M}|^2\rangle( D^0 \to \pi^- \pi^+ \pi^0 ) } \over
  {\langle|{\cal M}|^2\rangle( D^0 \to K^- \pi^+ \pi^0 ) }   } =  (6.68 \pm 0.04  \pm 0.08)\%
\label{eqn:Mratio1}
\end{equation}
\begin{equation}
{ {\langle|{\cal M}|^2\rangle( D^0 \to K^- K^+ \pi^0 ) } \over
  {\langle|{\cal M}|^2\rangle( D^0 \to K^- \pi^+ \pi^0 ) }   } = (4.53 \pm 0.06  \pm 0.08)\%
\label{eqn:Mratio2}
\end{equation}
\begin{equation}
{ {\langle|{\cal M}|^2\rangle( D^0 \to K^- K^+ \pi^0 ) } \over
  {\langle|{\cal M}|^2\rangle( D^0 \to \pi^- \pi^+ \pi^0 ) } } = (6.78 \pm 0.14  \pm 0.21)\%
\label{eqn:Mratio3}
\end{equation}
Belle
\begin{equation}
{ {\langle|{\cal M}|^2\rangle( D^0 \to \pi^- \pi^+ \pi^0 ) } \over
  {\langle|{\cal M}|^2\rangle( D^0 \to K^- \pi^+ \pi^0 ) }   } =  (6.13 \pm 0.06  \pm 0.19)\%
\label{eqn:Mratio4}
\end{equation}

To the extent that the differences in the matrix elements are only due to
Cabibbo-suppression at the quark level, the ratios of
the  matrix elements squared for singly Cabibbo-suppressed decays to  that
for the Cabibbo-favored decay should be approximately
$ \sin^2 \theta_C \approx 0.05 $ and the ratio of the  matrix elements squared
for the two singly Cabibbo-suppressed decays should be unity.
The deviations from this naive picture are less than 35\% for
these three-body decays.
In contrast, the corresponding ratios may be calculated for the
two-body decays $ D^0 \to \pi^- \pi^+ $, $ D^0 \to K^- \pi^+ $,
and $ D^0 \to K^- K^+ $.
Using the world average values for two-body branching
ratios~\cite{PDG},
the ratios of the  matrix elements squared for two-body Cabibbo-suppressed decays,
corresponding to \mbox{Eqs.~\ref{eqn:Mratio1}--\ref{eqn:Mratio4}},
are, respectively, $0.034 \pm 0.001$, $0.111 \pm 0.002$, and $3.53 \pm 0.12$.
Thus the naive Cabibbo-suppression model works well for three-body decays but not
so well for two-body decays.\\

\section{Amplitude analysis of $\mathbf{D}$ and $\mathbf{D_s}$ decays}
\begin{figure*}[t!]
\begin{center}
\includegraphics[width=17cm]{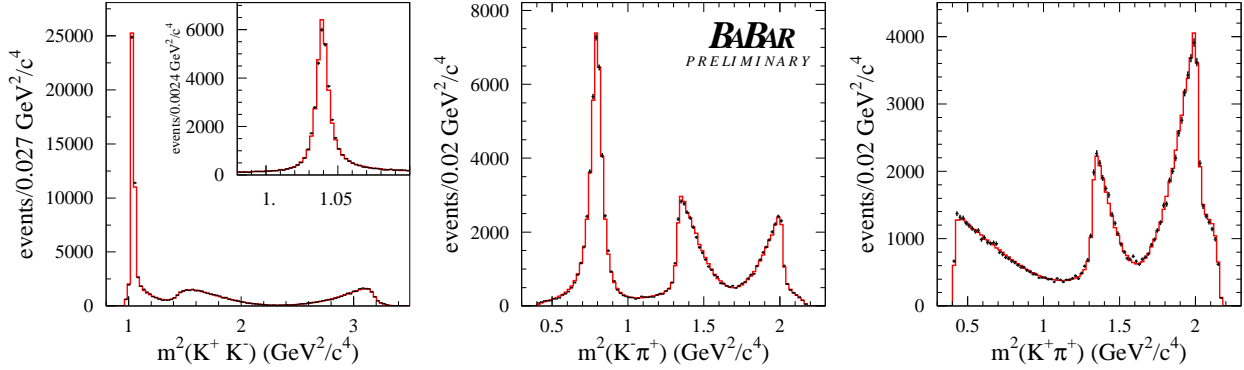}
\caption{The $D^+ \to K^+ K^- \pi^+$ Dalitz plot projections.
The data are represented by the points with error bars; the solid histograms are the projections of the fit described in the text. The inset shows an expanded view of the $\phi(1020)$ region.}
\label{fig:fig2}
\end{center}
\end{figure*}
The Dalitz plot analysis is the most complete method of studying
the dynamics of three-body charm decays.
These decays are
expected to proceed through intermediate quasi-two-body modes~\cite{two} and experimentally
this is the observed pattern.
Dalitz plot analyses can also provide new information on the resonances that
contribute to observed three-body final states.
In this kind of analysis the complex quantum mechanical amplitude $f$ is a coherent sum of all relevant
quasi-two-body $\Dz\to(r\to AB)C$ isobar model~\cite{isobar} resonances,
$f = \sum_r a_r e^{i\phi_r} A_r(s)$. Here $s=m_{AB}^2$, and $A_r$ is the
resonance amplitude.
The isobar model is expected to fail when there are large and overlapping resonances. In such case the $\pi\pi$ $S$-wave is often parameterized through a K-matrix formalism~\cite{ref:Kmatrix,ref:aitchison}.

\subsection{$\mathbf{D_s^+ \to K^+ K^- \pi^+}$ Dalitz plot analysis}

\babar\ collaboration reports the study of the three-body
\Ds meson decays to \Kp\Km\pip
 and in particular the measurement
of the branching fractions
$\frac{\BR(\Ds \to \phi \pip)}{\BR(\Ds \to \Kp \Km \pip)}$
and $\frac{\BR(\Ds \to \Kstarzb \Km)}{\BR(\Ds \to \Kp \Km \pip)}$.
The decay $\Ds \to \phi \pip$
is frequently used in particle physics as the \Ds reference decay mode.
The improvement in the measurements of these ratios is therefore important
because it allows the \Ds $\to$  \Kp\Km\pip to be used as reference.

A sample of 101k events with a purity of 95\% is selected by a likelihood function using vertex separation and $p^*$, the momentum of $D_s^+$ in CM system.
 A 66\% of this final sample consists of $D_s^+$'s originating from
 $D^*_s(2112)^+ \to D^+_s \gamma$ decay where
the variable
$$\Delta m = m(\Kp \Km \pip \gamma) - m(\Kp \Km \pip)$$
is required to be
 within $\pm 2\sigma$ of the PDG value~\cite{PDG}.

The selection efficiency is
determined from a sample of Monte-Carlo events in which the \Ds
decay is generated
according to phase-space.

An unbinned maximum likelihood fit is performed
in order to use the distribution of events
in the Dalitz plot to determine the relative amplitudes and phases
of intermediate resonant and non-resonant states.

The best fit results showing fractions,
are summarized in Tab.~\ref{tab:res}. The decay results to be dominated by $K^*(892)$ and $\phi$. Their branching ratio are:
$$\frac{\BR(\Ds \to \phi \pip)}{\BR(\Ds \to \Kp \Km \pip)}=0.379 \pm 0.002 \pm 0.018$$
and
$$\frac{\BR(\Ds \to \bar K^*(892)^0 \Kp)}{\BR(\Ds \to \Kp \Km \pip)}= 0.487 \pm
0.002 \pm 0.016$$
where errors are statistic and systematic respectively. These measurements are much more precise than the previous ones, based on a Dalitz plot analysis of only 700 events~\cite{e687}.

A $f_0(890)$ contribution is large but it is affected by large systematic errors as well due to uncertainness on $f_0(980)$ and $f_0(1370)$ parameters.
The Dalitz plot projections together with the fit results are shown in Fig.~\ref{fig:fig2}.

\begin{table}[!htb]
\centering
\caption{Fit fractions of a Dalitz plot fit of $D_s^+ \to K^+ K^- \pi^+$ decay. Errors are  statistical and systematic respectively.}

\begin{tabular}{l r@{}c@{}l}
\hline
Decay Mode                   & \multicolumn{3}{c}{Decay fraction(\%)}\\
\hline
\hline
$\bar K^*(892)^0 K^+$\phantom{{\LARGE A}} & $48.7 \, \pm \, $& $0.2$&$\, \pm \, 1.6$\\
$\phi(1020)\pi^+$            & $37.9 \, \pm \, $& $0.2$&$\, \pm \, 1.8$    \\
$f_0(980)\pi^+$              & $35 \, \pm \, $  & $1$&$\, \pm \, 14$       \\
$\bar K^*_0(1430)^0 K^+$     & $2.0 \, \pm \, $ & $0.2$&$\, \pm \, 3.3$    \\
$f_0(1710)\pi^+$             & $2.0  \, \pm \, $& $0.1$&$\, \pm \, 1.0$    \\
$f_0(1370)\pi^+$             & $6.3  \, \pm \, $& $0.6$&$\, \pm \, 4.8$    \\
$\bar K^*_2(1430)^0 K^+$     & $0.17 \, \pm \, $ & $0.05$&$\, \pm \, 0.3$  \\
$f_2(1270)\pi^+$             & $0.18 \, \pm \, $ & $0.03$&$\, \pm \, 0.4$  \\
\hline
\end{tabular}
\label{tab:res}
\end{table}
\begin{figure*}[t]
\centering
\includegraphics[width=55mm]{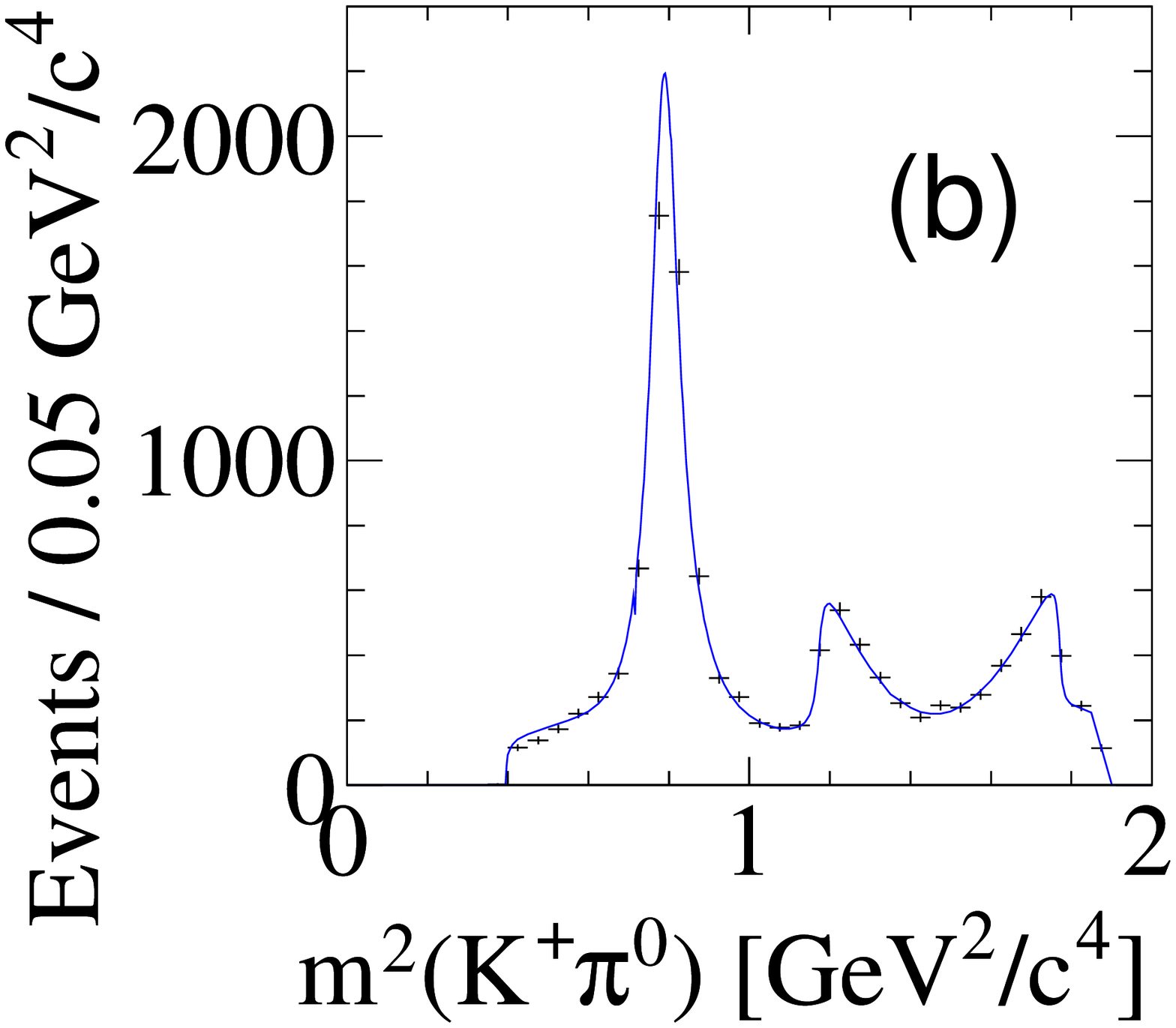}
\includegraphics[width=55mm]{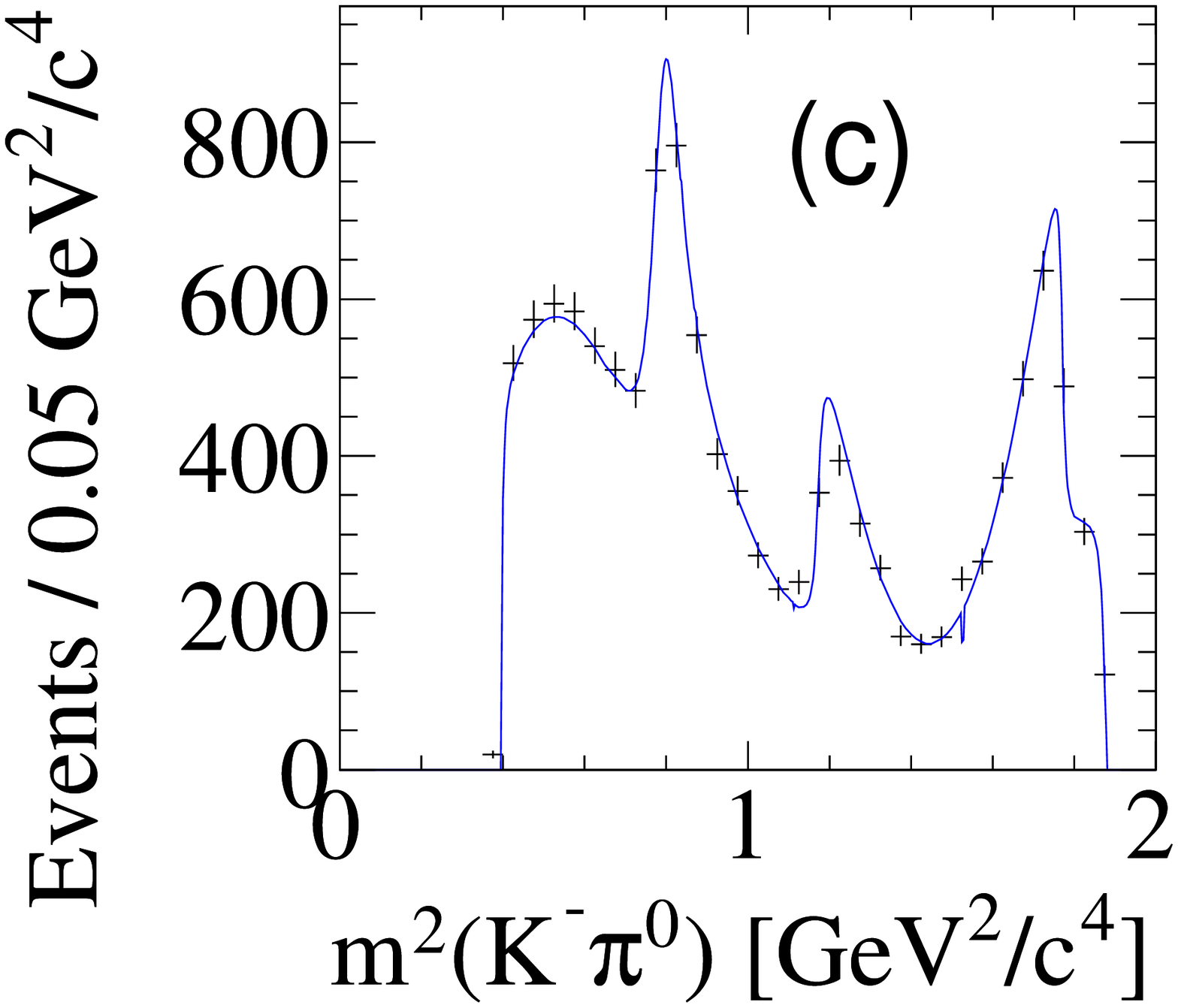}
\includegraphics[width=55mm]{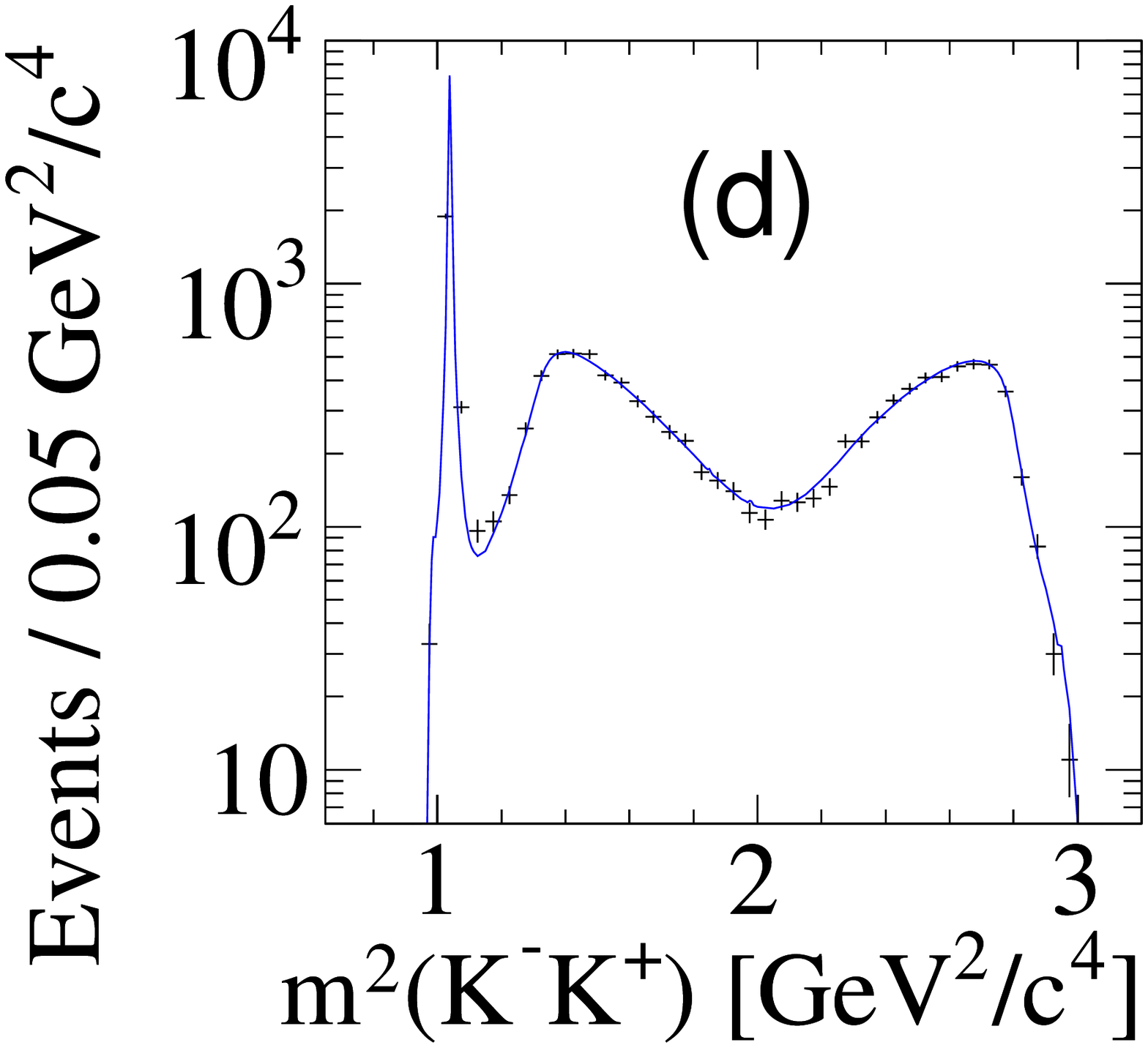}
\caption{Squared invariant mass projections of $D^0 \to K^- K^+ \piz$ Dalitz plot. The dots (with error bars, black) are data
points and the solid lines (blue) correspond to the isobar fit model.}
\label{fig:d0_kkpi}
\end{figure*}
Further tests of the fit quality are performed using unnormalized $Y^0_L$ moment projections
onto the $K^+K^-$ and $K^- \pi^+$ axis as functions of the helicity angles $\theta_K$
and $\theta_{\pi}$.
For $K^+ K^-$, the angle $\theta_K$ is defined as the
angle between the $K^-$ for $D^+_s$ (or $K^+$ for $D^-_s$)
in the $K^+ K^-$ rest frame and
the $K^+ K^-$ direction in the $D^+_s$
rest frame. The $K^+ K^-$ mass distribution is then modified by weighting
by the spherical harmonic $Y_L^0(\cos \theta_K)$
(L=1--4). A similar procedure is followed for the $K^- \pi^+$ system.
The resulting $\left<Y^0_1 \right>$ distributions are shown in
Fig.~\ref{fig:fig3}.

\begin{figure}[!htb]
\begin{center}
\includegraphics[width=80mm]{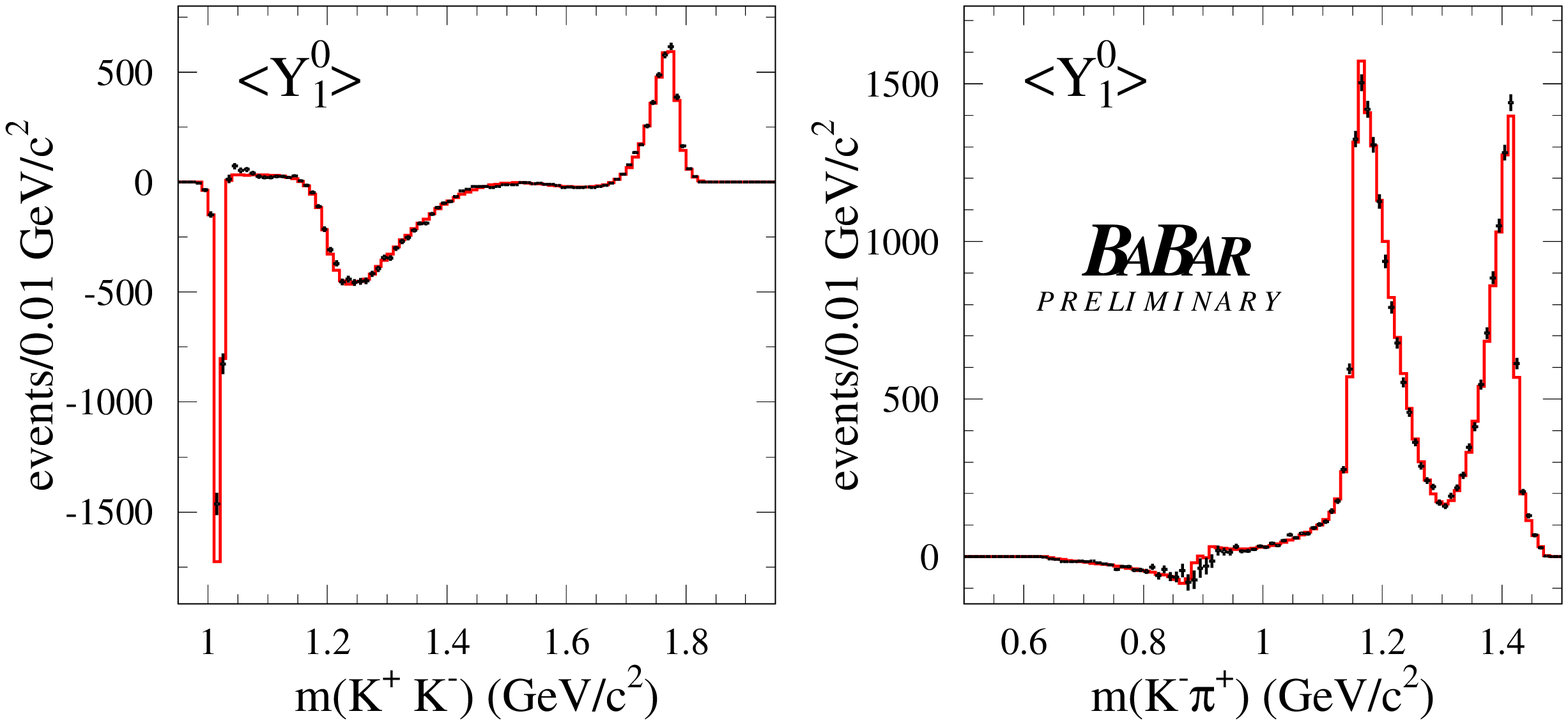}
\caption{The unnormalized spherical harmonic moments $\left <Y^0_1 \right >$
as a function of $K^+ K^-$ and  $K^- \pi^+$
effective masses. The data are presented with error bars, the solid
histograms represents the fit projections.
}
\label{fig:fig3}
\end{center}
\end{figure}

In order to interpret these distributions, one should recall the relationship between
$\left<Y^0_1 \right>$ moments and $S$- and
$P$-wave amplitudes~\cite{chung}:
\begin{equation}
\sqrt{4 \pi} \left<Y^0_1 \right> = 2 \mid S \mid \mid P \mid \cos \phi_{SP}
\label{eq:y01}
\end{equation}
Here $S$ and $P$ are proportional to the size of the $S$- and $P$-wave
contributions and $\phi_{SP}$ is their relative phase.  So $\left<Y^0_1 \right>$ results to be related to the $S$-$P$ interference.
Due to the presence of strong
reflections on the $K^+ K^-$ channel from the $K^- \pi^+$ channel (and vice versa), Eq.~\ref{eq:y01} is meaningful only in the threshold regions. Figure~\ref{fig:fig3} shows a large activity in the low $K^+ K^-$
mass distribution, suggesting the presence of a large $S$-wave contribution below the
$\phi(1020)$. The $\left<Y^0_1 \right>$ distribution along the $K^- \pi^+$ projection,
on the other hand,
has a very small activity in the $\bar K^*(892)^0$, suggesting a small $K \pi$ $S$-wave contribution.

\subsection{$\mathbf{D^0 \to K^+ K^- \pi^0}$ Dalitz plot analysis}

 The $K^{\pm}\piz$ systems~\cite{:2007dc} from the decay
\Dzkkpz can provide information on the $K\pi$ \textit{S-}wave amplitude in the mass range 0.6--1.4 \gevcc, and hence on the
possible existence of the $\kappa(800)$, reported to date only in the neutral
state ($\kappa^0 \to K^- \pi^+$)~\cite{kappa}. If the $\kappa$ has isospin
$1/2$, it should be observable also in the charged states. Results of the
present analysis can also be an input for extracting the $C\!P$-violating phase
$\gamma$~\cite{abi, myGamma}.

\Dz from \Dzb are identified by reconstructing
the decays $D^{*+}\to\Dz\pi^{+}$ and $D^{*-}\to\Dzb\pi^{-}$.
The signal efficiency is estimated for each event as a function of its
position in the Dalitz plot using simulated \Dzkkpz\ events from
$c\overline c$ decays, generated uniformly in the available phase space.

\indent For \Dz\ decays to $K^{\pm}\piz$ \textit{S-}wave states, three
amplitude models are considered: the LASS amplitude for
$K^-\pi^+\to K^-\pi^+$ elastic scattering~\cite{LASS},
 the E-791 results for the $K^-\pi^+$ \textit{S-}wave
amplitude from an energy-independent partial-wave analysis in the
decay $D^+\to K^-\pi^+\pi^+$~\cite{brian} and a coherent
sum of a uniform nonresonant term, and Breit-Wigner terms for the
$\kappa(800)$ and $K^*_0(1430)$ resonances.

\begin{figure*}[t]
\centering
\includegraphics[width=55mm]{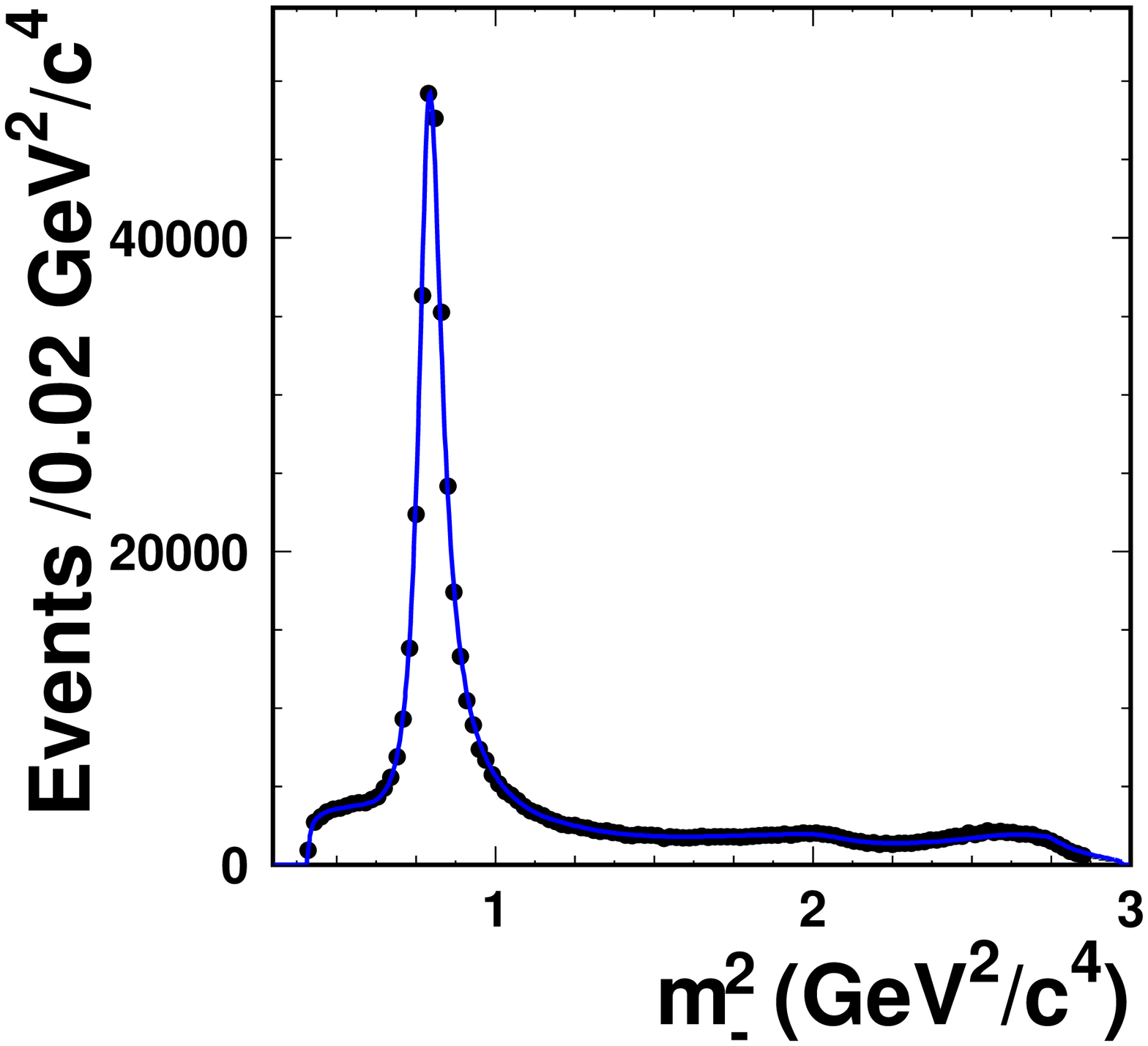}
\includegraphics[width=55mm]{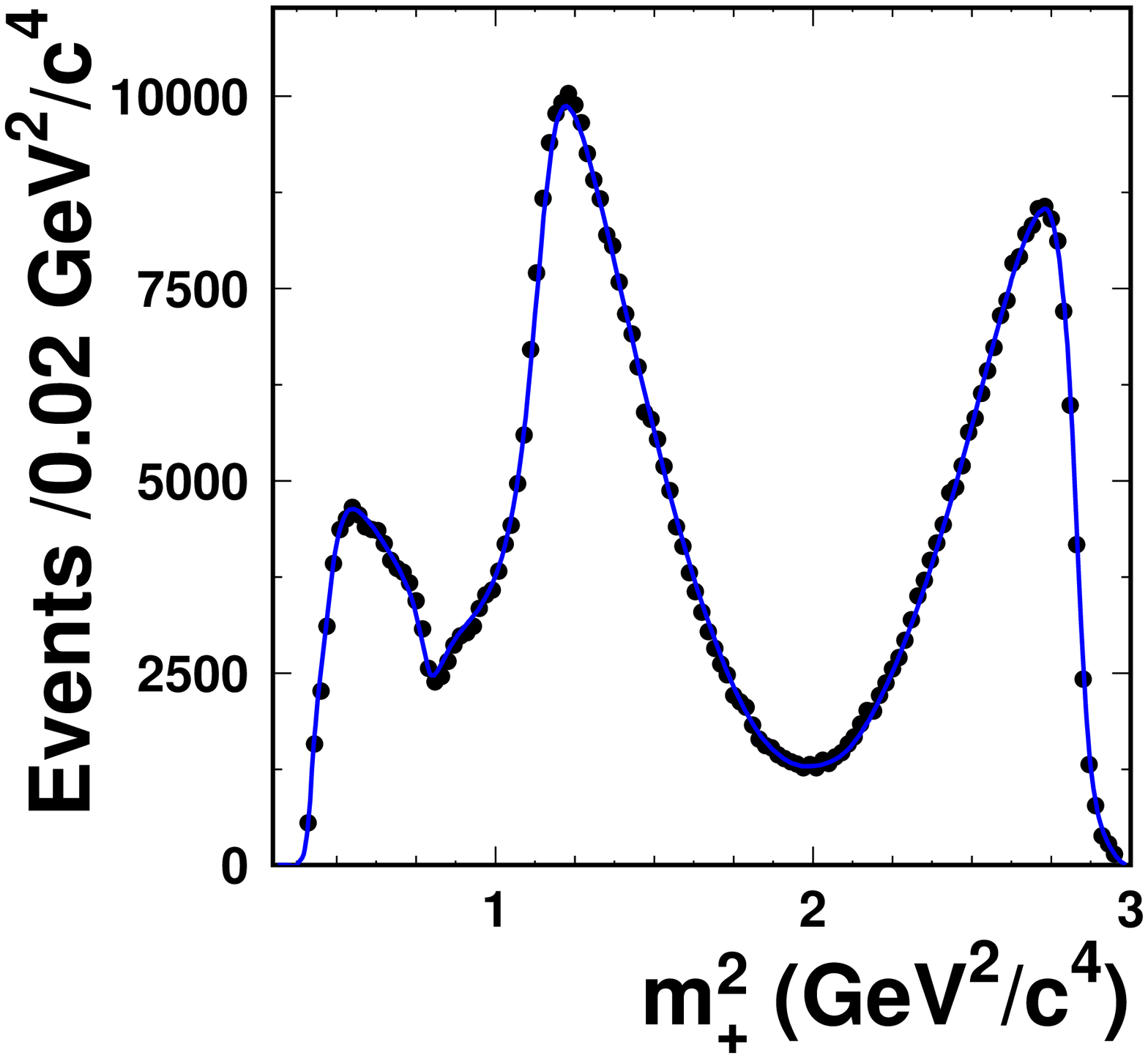}
\includegraphics[width=55mm]{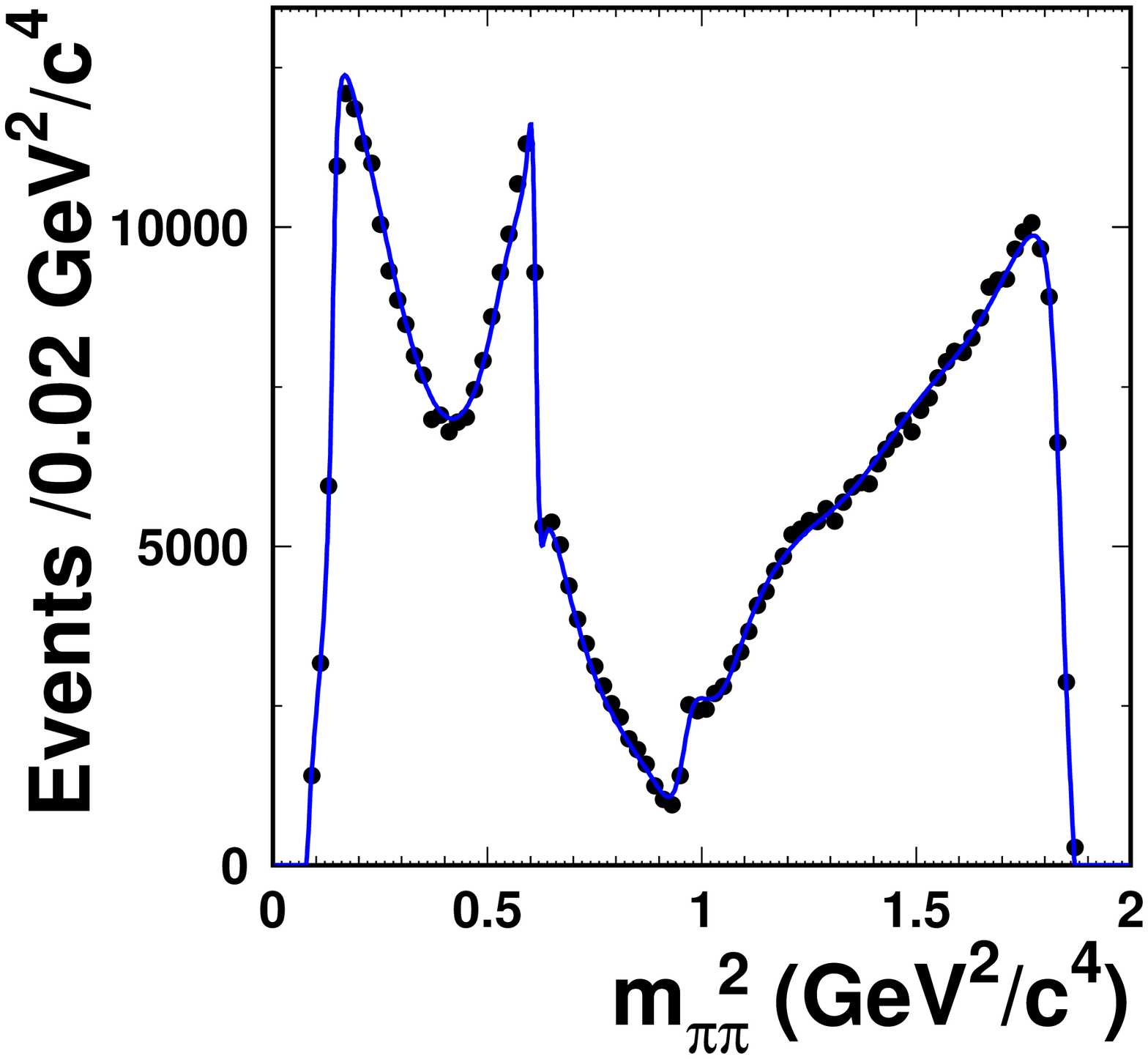}
\caption{Dalitz plot distribution and the
projections for data (points with error bars) and the fit result
(curve). Here, $m^2_\pm$ corresponds to $m^{2}(K_S^0 \pi^\pm)$ for
$\Dz$ decays and to $m^{2}(K_S^0\pi^\mp)$
 for $\Dzb$ decays(Belle collaboration).}\label{dfit_final}
\end{figure*}

The results of an unbinned maximum likelihood are shown in Fig.~\ref{fig:d0_kkpi}. While the measured
fit fraction(Tab.~\ref{tab:result}) for $\Dz\to K^{*+}K^-$ agrees well with a phenomenological
prediction~\cite{theory} based on a large SU(3) symmetry breaking, the
corresponding results for $\Dz\to K^{*-}K^+$ and the color-suppressed
$\Dz\to\phi\pi^0$ decays differ significantly from the predicted values.
The $K\pi$ \textit{S-}wave amplitude is consistent with that from
the LASS analysis, throughout the available mass range. 
The $K^-K^+$
\textit{S-}wave amplitude, parameterized as either $f_0(980)$ or $a_0(980)^0$,
is required. No higher mass $f_0$ states are found to
contribute significantly.

\begin{table}[htbp]
\caption{The results obtained from the $D^0 \to K^- K^+ \piz$ Dalitz plot
fit. The errors are statistical and systematic,
respectively. The $a_0(980)$ contribution, when it is included in
place of the $f_0(980)$, is shown in square brackets. LASS amplitude
is used to describe the $K\pi$ \textit{S-}wave states.}
\label{tab:result}
\begin{tabular}{l r@{}c@{}l}
\hline
State                      &  \multicolumn{3}{c}{Decay fraction(\%)}\\
\hline
\hline
$K^*(892)^{+}$             & $45.2 \, \pm$&$0.8$&$\pm \, 0.6$ \cr
$K^*(1410)^{+}$            &  $3.7 \, \pm$&$1.1$&$\pm \, 1.1$ \cr
$K^+\piz(\textit{S})$     &  $16.3 \, \pm$&$3.4$&$\pm \, 2.1$ \cr
$\phi(1020)$              &  $19.3 \, \pm$&$0.6$&$\pm \, 0.4$ \cr
$f_0(980)$                &  $6.7 \, \pm$&$1.4$&$\pm \, 1.2$  \cr
$\left[a_0(980)^0\right]$ &  [$6.0 \, \pm$&$1.8$&$\pm \, 1.2$] \cr
$f_2'(1525)$              & $0.08 \, \pm$&$0.04$&$\pm \, 0.05$  \cr
$K^*(892)^{-}$            &  $16.0 \, \pm$&$0.8$&$\pm \, 0.6$ \cr
$K^*(1410)^{-}$           &  $4.8  \, \pm$&$1.8$&$\pm \, 1.2$  \cr
$K^-\piz(\textit{S})$     & $2.7 \, \pm$&$1.4$&$\pm \, 0.8$ \cr
\hline
\end{tabular}
\end{table}

\indent Neglecting $C\!P$ violation, the strong phase difference, $\delta_D$,
between the \Dzb and \Dz decays to $K^*(892)^{+}K^-$ state and their amplitude
ratio, $r_D$, are given by
$$r_D e^{i\delta_D} = \frac{a_{\Dz\to K^{*-}K^+}}{a_{\Dz\to K^{*+}K^-}}
{ } e^{i(\delta_{K^{*-}K^+}{ } - { } \delta_{K^{*+}K^-})}.$$
\babar\ finds $\delta_D$ =
$-35.5^\circ \pm 1.9^\circ$ (stat) $\pm 2.2^\circ$ (syst) and $r_D$ = 0.599
$\pm$ 0.013 (stat) $\pm$ 0.011 (syst). These results are consistent with the
previous measurements~\cite{cleo}, $\delta_D$ = $-28^\circ\pm 8^\circ$ (stat)
$\pm 11^\circ$ (syst) and $r_D$ = 0.52 $\pm$ 0.05 (stat) $\pm$ 0.04 (syst).

\subsection{$\mathbf{D^0 \to K^0_S \pi^+ \pi^-}$ Dalitz plot analysis}

Recently,
evidence for $D^0$-$\Dzb$ mixing has been found in $D^0\ra
K^+K^-/\pip\pim$~\cite{y_cp} and $D^0\ra K^+\pi^-$~\cite{kpi_BaBar}
decays. It is important to measure $D^0$-$\Dzb$ mixing in other decay
modes and to search for $CP$-violating effects in order to
determine whether physics contributions outside the SM are present.
Belle~\cite{Abe:2007rd} collaboration reports a measurement of $D^0$-$\Dzb$ mixing studying $D^0 \to K^0_S \pi^+ \pi^-$ decay. The relevance of this decay is enhanced by its role in determining the angle $\gamma\equiv\arg{\left[-V_{ud}^{}V_{ub}^{*}/V_{cd}^{}V_{cb}^{*}\,\right]}$ of the Unitarity Triangle.
In fact, various methods~\cite{ref:DKDalitz} have been proposed to
extract $\gamma$ using
$\Bm \to \Dztilde \Km$ decays, all exploiting the
interference between the color allowed $\Bm \to \Dz \Km$ ($\propto
V_{cb}$) and the color suppressed $\Bm \to \Dzb \Km$ ($\propto
V_{ub}$) transitions, when the \Dz and \Dzb are reconstructed
in a common final state. The symbol \Dztilde indicates either a \Dz or a \Dzb meson.
Among the \Dztilde decay modes studied so far the $\KS \pim \pip$ channel is
the one with the highest sensitivity to $\gamma$ because of the best
overall combination of branching ratio magnitude, $\Dz-\Dzb$
interference and background level.
\babar\ collaboration reports a measurement of the angle $\gamma$  by studying the Dalitz plot of $D^0 \to K^0_S \pi^+ \pi^-$~\cite{Aubert:2006am}. In order to estimate the systematic errors due to model, \babar\ reports a Dalitz plot analysis where the $\pi\pi$ $S$-wave is parameterized by a K-matrix model~\cite{Aubert:2005yj}.

The results of these analyses are summarized in Tab.~\ref{tab:d0_kspipi}.
The decay is dominated by the $K^{*}(892)^-$ and $\rho(770)$ contribution. In order to improve the quality of fits, doubly Cabibbo suppressed $K^*$ contributions and two Breit-Wigner amplitudes $\sigma_1$ and $\sigma_2$ (whose masses and widths are float parameters) are included. $\sigma_1$ and $\sigma_2$ take in account the poor knowledge of $S$-wave in the low mass spectrum and $f_0(980)$ parameters. The K-matrix model overcomes this problem describing the $\pi\pi$ $S$-wave at all.

Figure~\ref{dfit_final} shows the results of unbinned maximum likelihood fit performed by Belle~\cite{Abe:2007rd}.

\begin{table}[htbp]
\caption{Summary of branching ratios of $\Dz \to \ks \pi^+ \pi^-$ Dalitz plot fits performed by Belle(Isobar Model) and \babar(Isobar and K-matrix Model).}
\label{tab:result2}
\begin{tabular}{|l|ccc|}
\cline{2-4}
\multicolumn{1}{c}{} & \multicolumn{1}{|c|}{Belle} & \multicolumn{2}{|c|}{BaBar} \\
\cline{2-4}
\multicolumn{1}{c}{} & \multicolumn{1}{|c|}{Isobar Model} & \multicolumn{1}{|c|}{Isobar Model} & \multicolumn{1}{|c|}{K-matrix Model} \\
\hline
State     & \multicolumn{3}{|c|}{Fit Fraction(\%)}\\
\hline
$K^*(892)^-$    & 62.27 & 58.1 & 58.9 \\
$K_0^*(1430)^-$ & 7.24  & 6.7  & 9.1  \\
$K_2^*(1430)^-$ & 1.33  & 3.6  & 3.1   \\
$K^*(1410)^-$   & 0.48  & 0.1  & 0.2   \\
$K^*(1680)^-$   & 0.02  & 0.6  & 1.4   \\
\hline
$K^*(892)^+$    & 0.54  & 0.5  & 0.7   \\
$K_0^*(1430)^+$ & 0.47  & 0.0  & 0.2   \\
$K_2^*(1430)^+$ & 0.13  & 0.1  & 0.0   \\
$K^*(1410)^+$   & 0.13  & ---  & ---   \\
$K^*(1680)^+$   & 0.04  & ---  & ---   \\
\hline
$\rho(770)$     & 21.11 & 21.6 & 22.3  \\
$\omega(782)$   & 0.63  & 0.7  & 0.6   \\
$f_2(1270)$     & 1.8   & 2.1  & 2.7   \\
$\rho(1450)$    & 0.24  & 0.1  & 0.3   \\
\hline
$f_0(980)$      & 4.52  & 6.4  &       \\
$f_0(1370)$     & 1.62  & 2.0  & S-wave\\
$\sigma_1$      & 9.14  & 7.6  & 16.2  \\
$\sigma_2$      & 0.88  & 0.9  &       \\
NR              & 6.15  & 8.5  &       \\
\hline
\end{tabular}
\label{tab:d0_kspipi}
\end{table}

\begin{acknowledgments}
We are grateful for the 
extraordinary contributions of our \pep2\ colleagues in
achieving the excellent luminosity and machine conditions
that have made this work possible.
The success of this project also relies critically on the 
expertise and dedication of the computing organizations that 
support \babar.
The collaborating institutions wish to thank 
SLAC for its support and the kind hospitality extended to them. 
This work is supported by the
US Department of Energy
and National Science Foundation, the
Natural Sciences and Engineering Research Council (Canada),
the Commissariat \`a l'Energie Atomique and
Institut National de Physique Nucl\'eaire et de Physique des Particules
(France), the
Bundesministerium f\"ur Bildung und Forschung and
Deutsche Forschungsgemeinschaft
(Germany), the
Istituto Nazionale di Fisica Nucleare (Italy),
the Foundation for Fundamental Research on Matter (The Netherlands),
the Research Council of Norway, the
Ministry of Education and Science of the Russian Federation, 
Ministerio de Educaci\'on y Ciencia (Spain), and the
Science and Technology Facilities Council (United Kingdom).
Individuals have received support from 
the Marie-Curie IEF program (European Union) and
the A. P. Sloan Foundation.

\end{acknowledgments}

\bigskip 

\end{document}